\documentclass[%
 reprint,
 amsmath,amssymb,aps,
 prx,superscriptaddress,
 longbibliography
]{revtex4-2}
\usepackage[utf8]{inputenc}
\usepackage{graphicx}
\usepackage{amsmath}
\usepackage{dcolumn}
\usepackage{bm}
\usepackage{xcolor}
\usepackage{siunitx}
\usepackage{times}
\usepackage[hidelinks]{hyperref}
\hypersetup{
  colorlinks   = true, 
  urlcolor     = blue, 
  linkcolor    = blue, 
  citecolor   = blue 
}
\usepackage{amssymb}
\usepackage{verbatim}
\DeclareUnicodeCharacter{03B3}{$\gamma$}
\DeclareUnicodeCharacter{03B4}{$\delta$}
\DeclareUnicodeCharacter{2212}{$-$}
\DeclareUnicodeCharacter{226A}{$\ll$}
\DeclareUnicodeCharacter{03BE}{$\xi$}

\usepackage{booktabs}

\DeclareSIUnit\bar{bar}

\begin{document}

\title{Identification of multiple-flux-quanta vortices by core states in the two-band superconductor Pb 
}

\author{Thomas Gozlinski}
\email{thomas.gozlinski@kit.edu}
\thanks{These authors contributed equally to this work.}
\affiliation{Physikalisches Institut, Karlsruhe Institute of Technology,
Wolfgang-Gaede-Str.1, 76131 Karlsruhe, Germany}
\author{Qili Li}
\thanks{These authors contributed equally to this work.}
\affiliation{Physikalisches Institut, Karlsruhe Institute of Technology,
Wolfgang-Gaede-Str.1, 76131 Karlsruhe, Germany}
\author{Rolf Heid}
\affiliation{Institute for Quantum Materials and Technologies, Karlsruhe Institute of Technology, Hermann-von-Helmholtz-Platz 1, 76344 Eggenstein-Leopoldshafen, Germany}
\author{Ryohei Nemoto}
\affiliation{Department of Materials Science, Chiba University,
1-33 Yayoi-cho, Inage-ku, Chiba 263-8522, Japan}
\author{Roland Willa}
\affiliation{Institute for Theory of Condensed Matter, Karlsruhe Institute of Technology,
Wolfgang-Gaede-Str.1, 76131 Karlsruhe, Germany}
\author{Toyo Kazu Yamada}
\affiliation{Department of Materials Science, Chiba University,
1-33 Yayoi-cho, Inage-ku, Chiba 263-8522, Japan}
\affiliation{Molecular Chirality Research Centre, Chiba University,1-33 Yayoi-cho, Inage-ku, Chiba 263-8522, Japan}
\author{Jörg Schmalian}
\affiliation{Institute for Quantum Materials and Technologies, Karlsruhe Institute of Technology, Hermann-von-Helmholtz-Platz 1, 76344 Eggenstein-Leopoldshafen, Germany}
\affiliation{Institute for Theory of Condensed Matter, Karlsruhe Institute of Technology,
Wolfgang-Gaede-Str.1, 76131 Karlsruhe, Germany}
\author{Wulf Wulfhekel}
\affiliation{Physikalisches Institut, Karlsruhe Institute of Technology,
Wolfgang-Gaede-Str.1, 76131 Karlsruhe, Germany}
\affiliation{Institute for Quantum Materials and Technologies, Karlsruhe Institute of Technology, Hermann-von-Helmholtz-Platz 1, 76344 Eggenstein-Leopoldshafen, Germany}


\date{\today}

\begin{abstract}

Superconductors are of type I or II depending on whether they form an Abrikosov vortex lattice. Although bulk lead (Pb) is classified as a prototypical type-I superconductor, we observe single-flux-quantum and multiple-flux-quanta vortices in the intermediate state using $\SI{}{\milli\kelvin}$ scanning tunneling microscopy. We show that the winding number of individual vortices can be determined from  the real space wave function of its Caroli-de Gennes-Matricon bound states. This generalizes the index theorem put forward by Volovik for isotropic electronic states to realistic electronic structures. In addition, the bound states due to the two superconducting bands of Pb can be separately detected.  This yields strong evidence for low inter-band coupling and an independent closure of the gaps inside vortices.

\end{abstract}


\maketitle 




\section{\label{sec:level1}Introduction}
The classical solutions in Ginzburg-Landau theory (GL) allow a thermodynamic classification of superconductors into type-I and type-II. Decisive for their behaviour in magnetic field is the interface energy between the superconducting and normal phase driven by the ratio of the London magnetic penetration depth $\lambda_\mathrm{L}$ and the superconducting coherence length $\xi$. For Ginzburg-Landau parameters $\kappa=\lambda_\mathrm{L}/\xi <1/\sqrt{2}$, type-I behaviour and for $>1/\sqrt{2}$ type-II behaviour was predicted \cite{ginzburg_theory_1950}. A type-I superconductor is characterized by a positive interface energy and an attractive vortex-vortex interaction favouring an intermediate state with large normal domains \cite{landau_theory_1937}. A type-II superconductor is characterized by a negative interface energy and a repulsive vortex-vortex interaction that favours an Abrikosov lattice of single-flux-quantum vortices in the mixed phase \cite{Abrikosov1957}. 

Flux quantization $\Phi = n \Phi_0$ in units of the flux quantum  $\Phi_{0} = h/2e$ is one of the most fundamental traits of the underlying off-diagonal long-range order of the superconducting condensate \cite{yang_concept_1962}. $n$ is the integer winding number of the vortex. This number of confined flux quanta is expected to affect the size and shape of the vortex. More quantitatively,  Volovik demonstrated for superconductors with isotropic Fermi surfaces that one can determine $n$ from an index theorem, provided one can measure the wave function of the quasiparticle bound states that form in the vortex core \cite{volovik_localized_1991}. The vorticity is particularly ambiguous, and hence interesting near 
the Bogomol'nyi point, $\kappa=1/\sqrt{2}$. Then, vortices  near $T_c$  behave like non-interacting particles and the vortex configuration is infinitely degenerate \cite{bogomolnyi_stability_1976, jacobs_interaction_1979, weinberg_multivortex_1979, taubes_arbitraryn-vortex_1980} such that multiple-flux-quanta (or giant) vortices  may be stabilized \cite{efanov_exact_1997}. 
This degeneracy is lifted below $T_c$ and a transitional region, in which the superconductor cannot be categorized into either of the types described above, emerges \cite{lukyanchuk_theory_2001}. When leaving the Ginzburg-Landau limit towards lower temperatures and especially, when considering multiple band superconductors, microscopic interactions are predicted to become increasingly important: the vortex-vortex interaction energy can become non-monotonic in distance through the existence of multiple, distinct superconducting coherence lengths $\xi_i$ \cite{moshchalkov_type-15_2009, babaev_type-15_2012} and topological hysteresis due to transitions between flux tubes and laminar pattern influences the flux patterns of the intermediate state \cite{prozorov_topological_2005}. 

In the  past, the crucial parameter $\kappa$ has been tuned toward the type-II regime by either an increase of $\lambda_\mathrm{L}$ or reduction of the effective $\xi$, i.e. by using thin films below a critical thickness \cite{cody_magnetic_1968, dolan_critical_1973}, by incorporation of impurities (for relevant experiments see Ref.~\cite{lukyanchuk_theory_2001} and references within) or by interface scattering, e.g. Pb/Si(111) \cite{cren_ultimate_2009}. This approach, however, has the drawback that the quasiparticle bound states in the vortex core are considerably smeared out such that the index theorem cannot be applied \cite{Silaev2013,yamane_impurity_2013,masaki_impurity_2015,cren_ultimate_2009,cren_vortex_2011}. 

Here, we take the alternative approach and study the two-band superconductor Pb in the clean limit in form of a bulk single crystal Pb(111) at $\SI{45}{\milli\kelvin}$. In this respect, the multiple band superconductor Pb, which is closest to the Bogomol'nyi point of all elemental superconductors, is a good candidate to study the transitional phase at temperatures well below $T_c$. We map the quasiparticle bound states for the two superconducting gaps of Pb with high energy and spatial resolution and use the index theorem including realistic band structures to determine the winding number of single- and multiple-flux-quanta vortices. Ultimately, we show that our observations allow to investigate the inter-band coupling of the two gaps. 

\section{Results and discussions}
\begin{figure*}[t]
    \centering
    \includegraphics[width=\textwidth]{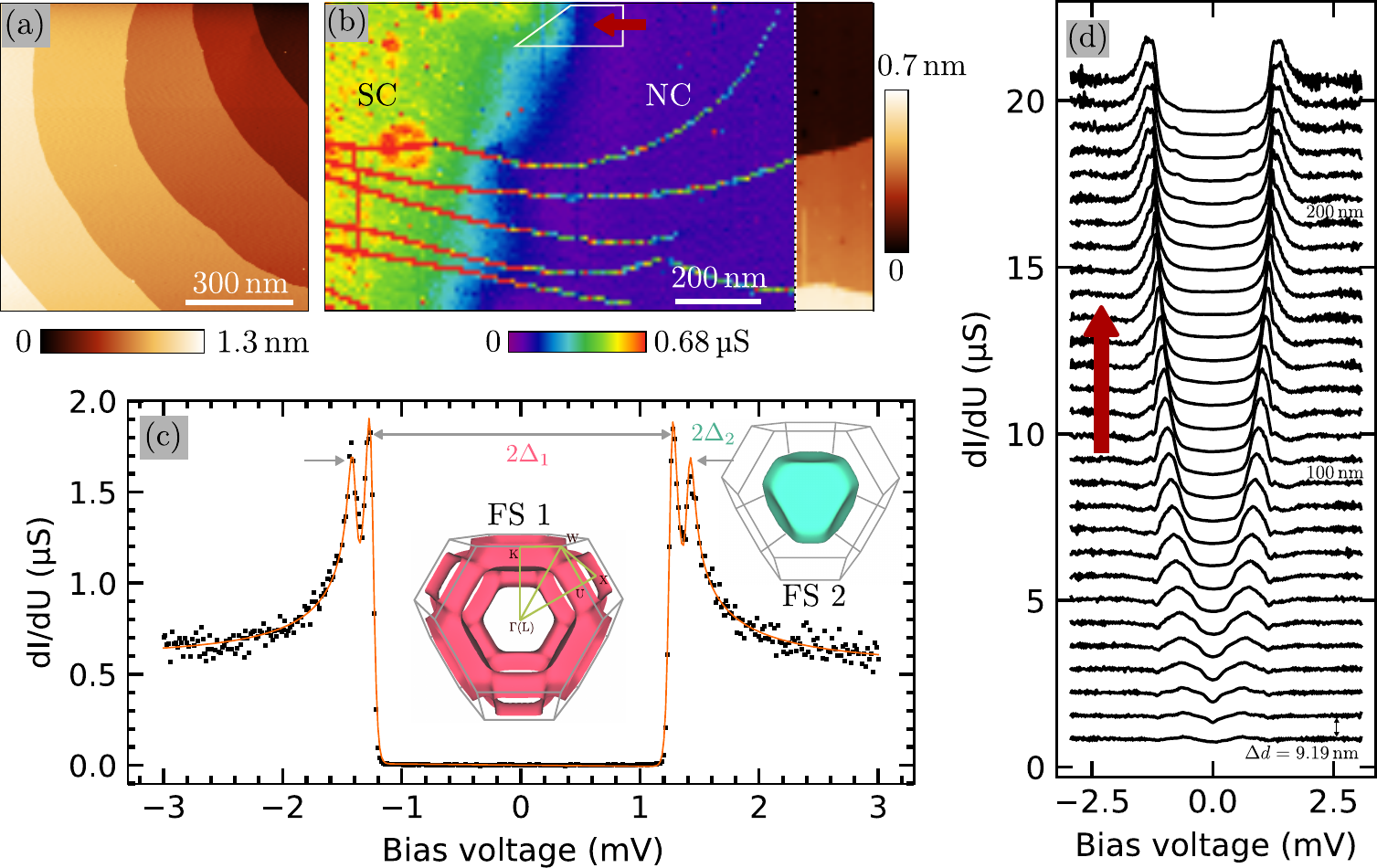}
    \caption{\textbf{Superconducting properties and intermediate state.} (a) Topographic scan image of the Pb(111) surface after the cleaning procedure. (b) d$I$/d$U$ map at $U_t=\SI{1.3}{\milli e\volt}$ showing a typical domain wall in the intermediate state at $B=\SI{23}{\milli\tesla}$. Right overlay: Corresponding topographic scan image. (c) Differential conductance in the superconducting state (or superconducting domain) (black points) including a two-gap BCS fit (orange) with $\Delta_1$ being the smaller and $\Delta_2$ being the larger gap. Inset: 3D models of Fermi surface sheets 1 and 2 (from Ref. \cite{choy_database_2000}) responsible for superconducting gaps $\Delta_1$ and $\Delta_2$ respectively. (d) d$I$/d$U$ spectra along a cross-section from normal conducting to superconducting domain. The area is marked in white in (a) and the profile direction is indicated by a red arrow. Individual spectra are offset with respect to each other by $\SI{0.7}{\micro S}$ The spectra were locally averaged over a straight part of the domain boundary and recorded in distance increments of $\Delta d=\SI{9.19}{\nano\metre}$.}
    \label{fig:intermediate_state}
\end{figure*}

\subsection{Superconducting gaps and intermediate state}
After several cycles of sputtering and annealing, we obtained a clean Pb(111) surface with wide terraces and monoatomic steps, as shown in Fig.~\ref{fig:intermediate_state}(a). Upon zero-field cooling the Pb(111) sample to $\SI{45}{\milli\kelvin}$ it enters its superconducting state below $T_c\approx \SI{7.2}{\kelvin}$ \cite{poole_handbook_2000}. Due to our low electronic temperature of less than $\SI{100}{\milli\kelvin}$ \cite{Balashov2018} we are able to resolve the two gaps \cite{Ruby2015} in the density of states by scanning tunneling spectroscopy, even with a normal conducting tip. Fig.~\ref{fig:intermediate_state}(c) shows the differential conductance in the superconducting state as black dots, including a temperature broadened two-gap BCS fit in orange. We determine the superconducting gaps to be $\Delta_1 = \SI{1.26(2)}{\milli e\volt}$ (smaller gap) and $\Delta_2 = \SI{1.40(2)}{\milli e\volt}$ (larger gap) in good agreement with previous measurement of the difference of the two gaps \cite{Ruby2015}. The intensity difference of the two coherence peaks has previously been attributed to the $k$-dependent tunneling matrix elements and the larger gap has been assigned to the tubular Fermi surface sheet \cite{Ruby2015}. This is in contrast to Bogoliubov-de Gennes (BdG) based Korringa-Kohn-Rostoker (KKR) calculations \cite{Saunderson2020}, which deduced an opposite band-to-gap assignment. As will be discussed below, our study of the quasi-particle bound  states in the vortices confirms the band-to-gap assignment of Saunderson et al. \cite{Saunderson2020}. We will from here on index the bands and Fermi surfaces according to their superconducting gap, i.e. the tubular Fermi surface responsible for $\Delta_1$ as Fermi surface 1 (FS 1) and the compact Fermi surface responsible for $\Delta_2$ as Fermi surface 2 (FS 2) (Fig.~\ref{fig:intermediate_state}(c)).

After applying a perpendicular magnetic field of $B = \SI{85}{\milli\tesla}$, which is above the critical field $\mu_0 H_c \sim \SI{80}{\milli\tesla}$ \cite{chanin_critical-field_1972}, magnetic flux enters the sample from the sides and completely destroys superconductivity. Upon decreasing the field again below $H_c$, the Landau intermediate state is reached. It is detected by recording the differential conductance at the coherence peak of $\Delta_2$ while ramping the field down. Once a jump to the superconducting state below the tip is detected, the ramp is stopped. This ensures that one typically finds both, superconducting and normal conducting, areas in the scan range of the STM of $1.4\times \SI{1.4}{\micro\meter^2}$. The intermediate state is characterized by large normal and superconducting domains. The shapes and sizes of these domains in the intermediate state of lead have been extensively studied by magneto-optical methods revealing the strong dependencies
on temperature, sample shape and magnetic protocol \cite{prozorov_topological_2005, prozorov_equilibrium_2007, prozorov_suprafroth_2008, prozorov_dynamic_2009}.

A typical domain wall in the intermediate state is shown in the d$I$/d$U$ map in Fig.~\ref{fig:intermediate_state}(b). At a tunneling bias of $U_t=\SI{1.3}{\milli\volt}$ the normal conducting domains show up as areas of low conductance (purple) and the superconducting domains as areas of high conductance (green/yellow). Note that atomic step edges of the surface cause a contrast in d$I$/d$U$ as typically found in STM experiments and is illustrated by the inset on the right.
A cross-sectional line scan across the domain wall, as in Fig.~\ref{fig:intermediate_state}(d), shows how both gaps change from zero to their maximum on the length scale of the coherence length. The local recovery of superconductivity agrees well with reported coherence lengths of $\xi=\SI{87}{\nano\metre}$ \cite{poole_handbook_2000}. For a detailed analysis of the coherence length $\xi_{1,2}$ of the two bands measured inside vortices, we refer to the next section.

\subsection{Single-flux-quantum vortices}

Inside the normal-core region of vortices, electronic bound states that lie within the superconducting gap are localized. These in-gap states were first theoretically studied by Caroli, de Gennes and Matricon (CdGM) in 1964 \cite{Caroli1964}.  
The CdGM states for isotropic bands can be characterized by their orbital angular momentum number $\mu$ and their energy. The energy spacing of the discrete CdGM states is of the order of $ \Delta^2/E_F$ and the discrete states thus form a quasicontinuum or branch of CdGM states for most superconducting materials. 
The CdGM states with low $\mu$ are confined closer to the vortex core than the ones with high $\mu$ thus leading to a one-to-one correspondence between the angular momentum and the real-space behavior of the bound-state wave function. The higher the angular momentum $\mu$ or energy $E(\mu)$, the  further away from the vortex center are the states localized \cite{gygi_self-consistent_1991}. 
In case of a $m$-flux quanta vortex, $m$ individual CdGM branches exist \cite{volovik_localized_1991, volovik_vortex_1993, volovik_topological_2019, Virtanen1999, Virtanen2000} leading to a topological index theorem: The number of zero-energy crossings of CdGM state branches with varying angular momentum is directly related to the vorticity \cite{volovik_localized_1991, volovik_topological_2019, volovik_vortex_1993}. This theorem translates to real space and the number of zero-energy crossings of CdGM state branches with varying radius from the vortex core is related to the vorticity. 

STM allows to measure the variation of the local density of states (LDOS) inside the vortex and thus to determine the winding number of the vortex using the index theorem. In 1989, Hess et al. experimentally confirmed that CdGM states exist in single-flux-quantum vortices by scanning tunneling microscopy (STM) \cite{Hess1989}, but 
for vortices with multiple flux quanta, although studied in thin films with electron holography \cite{Hasegawa1991} and scanning Hall probe microscopy \cite{Ge2013}, their predicted bound states still lack experimental verification \cite{Tanaka1993,Virtanen1999,Virtanen2000,Tanaka2002}.

\begin{figure*}[pt]
    \centering
    \includegraphics[width=\textwidth]{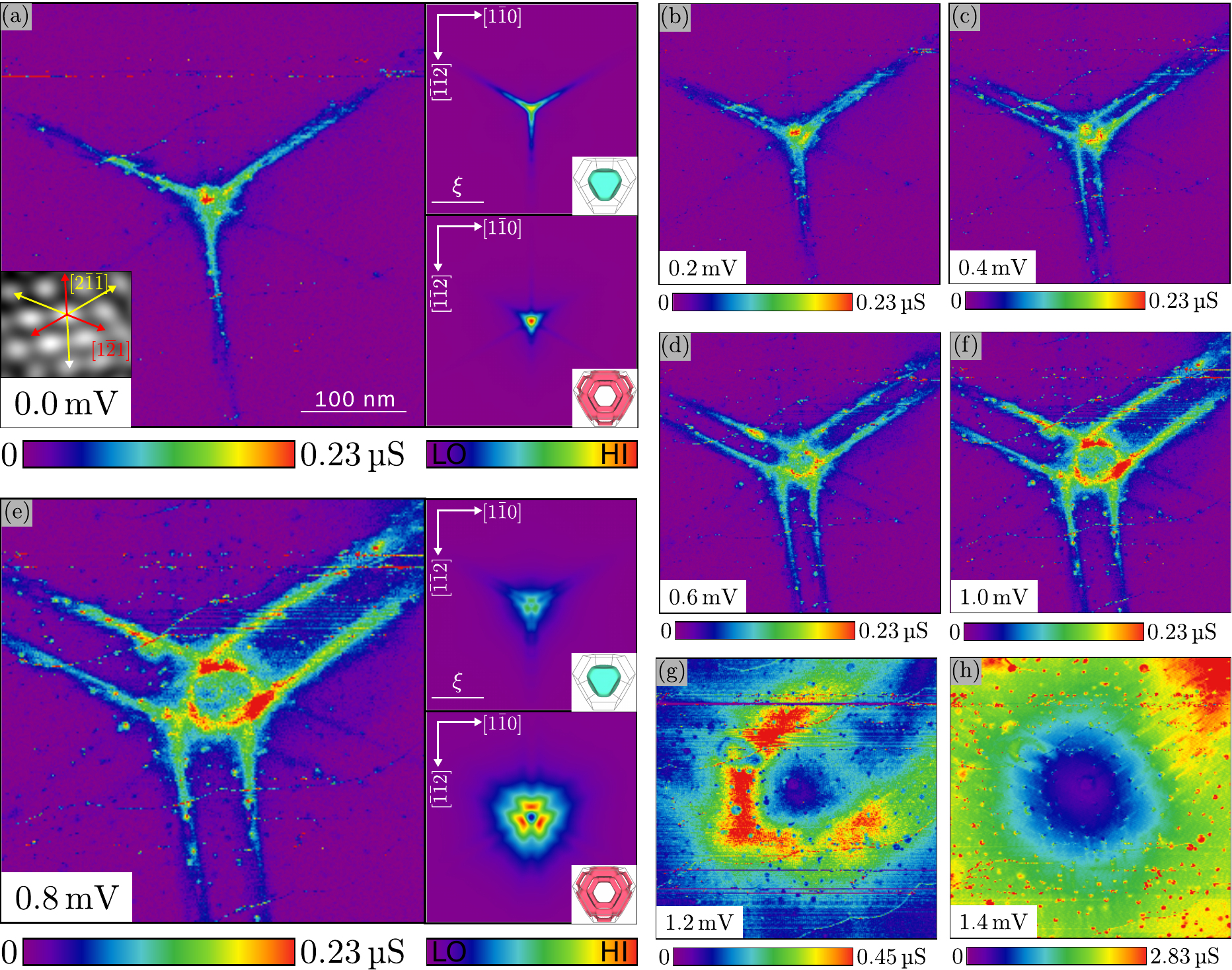}
    \caption{\textbf{Single-flux-quantum vortex signature.} (a-h) d$I/$d$U$ maps ($B=\SI{0}{\milli\tesla}$) of a single-flux-quantum vortex at different bias voltages displaying the quasiparticle density at the surface. Right panels in (a) and (e) show the simulated LDOS of the bands 1 and 2 for the respective energy. The inset in (a) shows a 2D-FFT filtered topographic image of the Pb(111) lattice. Bulk crystal directions (red/yellow arrows) have been determined from glide planes.}
    \label{fig:single_flux_vortex}
\end{figure*}

Using the detection method described in the previous section we are also able to find isolated, round normal conducting domains (appearance at $eU=\Delta_2$) of $\approx \SI{100}{\nano\metre}$ in diameter. An example is shown in Fig.~\ref{fig:single_flux_vortex}(h). As will be shown later, these are vortices in the superconductor with integer number of flux quanta.
The finding of such small normal conducting domains is surprising considering that the domain wall energy in type-I superconductors is positive and the system thus tries to maximize its domain size. Even more surprising is the fact, that we found this shape after ramping the field down from the critical field. Magneto-optical measurements of cylindrical shaped intermediate state lead samples at $\SI{4.5}{\kelvin}$ reveal that normal domains are only tubular upon increasing magnetic field; after ramping down from the critical field the preferred structure is laminar \cite{prozorov_equilibrium_2007}. A deciding factor for the  intermediate state domain structure on a microscopic scale could be the effect of flux branching \cite{landau_intermediate_1938, landau_theory_1943, conti_branched_2015}. Since the overall domain structure in our experiment, however, consists of domains of various shapes and sizes, we argue that this finding is only consistent with circumstances under which vortices are weakly-interacting and have a non-monotonous interaction energy causing hysteresis, i.e. a superconductor in the transitional phase (close to the Bogomol'nyi point) at temperatures well below $T_c$. 
The role of pinning of these vortices at bulk defects below the surface remains unclear. Repeating our magnetic protocol several times, the vortices may appear at similar positions suggestive of some kind of pinning. However, we also find the vortices to be mobile when varying the magnetic field (see next section).


To determine the amount of flux carried by the small normal domains, we record differential conductance (d$I$/d$U$) maps at sub-gap energies, which essentially show the LDOS of the CdGM states. At zero bias voltage, we find a threefold symmetric state in form of a star with a maximum in the star's centre (Fig.~\ref{fig:single_flux_vortex}(a)) inside the normal domain similar to star-shaped CdGM states seen by Hess et al. in $\mathrm{2}$H-NbSe$_\mathrm{2}$ \cite{Hess1990}. The quasiparticle density of states stretches over $\SI{100}{\nano\metre}$ in the $\langle 2\bar{1}\bar{1}\rangle$ directions (blue/green). 
Additionally, three weak arms (dark blue) along the $\langle 1\bar{2}1\rangle$ directions are visible. With increasing energy (independent of sign of bias voltage) the star's arms split into two with increasing splitting distance, while the central peak splits nearly isotropically to a ring shape (Fig.~\ref{fig:single_flux_vortex}(b-f)). For $E\lesssim \Delta_1$ the strong arms are still visible and the ring reaches its maximal size (Fig.~\ref{fig:single_flux_vortex}(g)). For $E\sim \Delta_2$ the vortex shows as a relatively round area of low conductance with $\approx \SI{100}{\nano\metre}$ in diameter (Fig.~\ref{fig:single_flux_vortex}(h)).  

For bands with anisotropic Fermi velocity, like in our case, the index theorem is not straight forward applicable since the radial symmetry is removed and a radial dependent measurement does not ensure crossing all diabolical points \cite{volovik_localized_1991, volovik_vortex_1993}. Diabolical points here mean the points where the semiclassical particle and hole spectrum meet. The degenerate gapless fermionic excitations or ``zero-modes'' at these points carry the topological charge (singularity in the phase) and owe their name to the double-conical (diabolo) dispersion in parameter space \cite{berry_diabolical_1997}.
Here, a realistic band structure needs to be considered. We, thus, simulated the quasiparticle trajectories inside a vortex carrying one flux quantum within the quasiclassical Eilenberger theory including the Fermi velocity of each band obtained from DFT calculations and compare the obtained LDOS maps to our experimental results (for details, see Methods).  
Right panels in Figure \ref{fig:single_flux_vortex}(a,e) display the solutions reproducing the star shape for the compact Fermi surface 2 and the ring like structure of the tubular Fermi surface 1. 
The simulations confirm that the observed states are the signature of a vortex in Pb(111) containing a single flux quantum.
The ring shaped states, related to the tubular Fermi surface (band 1) of nearly isotropic Fermi velocity, show the expected behaviour. At $U=0$, it creates a sharp maximum in the center of the vortex that splits into a ring of increasing diameter upon variation of the voltage away from $U=0$. For the compact Fermi surface (band 2) with large flat structures in the Fermi surface and anisotropic Fermi velocity, a star shaped structure is predicted whose arms split into two when going away from $U=0$. The observation of two sets of CdGM states in the experiment in good agreement with the semi-classical treatment of separate Fermi surfaces which already indicates that the coupling between the two Fermi surfaces is weak, i.e. the  rate of scattering between the two Fermi surfaces is significantly lower than the inverse time for the electron round trip around the vortex.  


Star shaped CdGM states have first been found by Hess et al. \cite{Hess1990} in 2H-NbSe$_2$, which due to the crystal symmetry have sixfold rotational symmetry. Pb crystallizes in the fcc structure and belongs to the point group $Fm\overline{3}m$. The electronic structure therefore only carries a threefold rotational symmetry about the [111] axis. Due to the discrete rotational symmetry, the angular momentum is not a good quantum number and the states mix. In general, the mixing results in states with different lateral confinement for different angles. At zero bias, the low angular momentum states carry the largest weight and lead to a maximum in the center of the vortex. At higher bias, states of larger angular momentum become increasingly important leading to a movement of the maxima away from the center, i.e. the star splits.  The large variation in quasiparticle localization depending on the angle from $\sim \SI{10}{\nano\metre}$ to over $\SI{100}{\nano\metre}$ agrees with the strong anisotropy of the 3D Fermi velocity in the compact band. 

\begin{figure*}[pt]
    \centering
    \includegraphics[width=\textwidth]{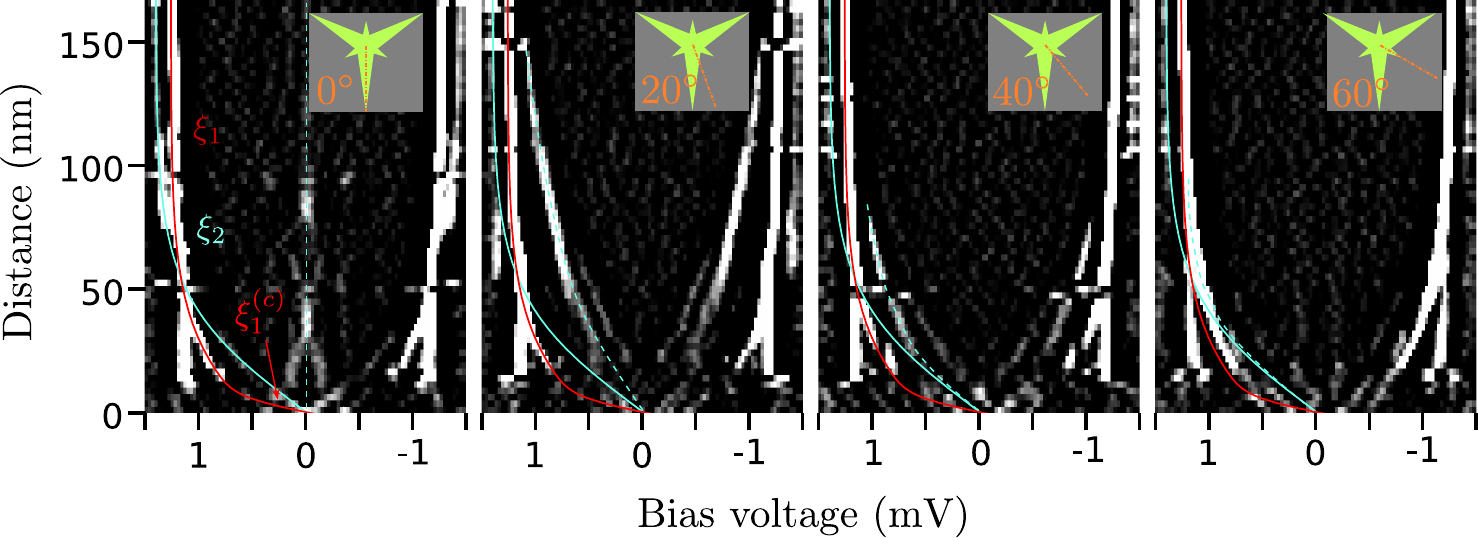}
    \caption{\textbf{Coherence lengths in the normal single-flux-quantum vortex.} Angle dependent radial measurement of the vortex core states from the vortex in Fig.~\ref{fig:single_flux_vortex}. Displayed is the second derivative of the differential conductance $-\partial^2 \sigma /\partial U^2$ in order to highlight maxima in the LDOS. The point distance of single spectra is $\Delta d=\SI{2.57}{\nano\metre}$. Marked are the opening of $\Delta_1$ (red line), the opening of $\Delta_2$ (cyan line) and the CdGM states of band 2 (cyan dashed line). The inset shows the direction of the y-axis.}
    \label{fig:angle_meas_normal}
\end{figure*}
In fact, star shaped vortices with long arms have been recently observed in the Abrikosov lattice of La(0001) \cite{kim_anisotropic_2021}, owing to the large anisotropy of the responsible band's in-plane Fermi velocity. 
In Pb with its two bands, one may expect that the individual gaps will close independently when approaching the vortex center. Fig.~\ref{fig:angle_meas_normal} displays the evolution of the gaps, i.e. the energy of the coherence peaks which appear as bright lines in the second derivative of d$I$/d$U$ with respect to $U$, as function of the distance from the vortex center and direction. For large distances from the center, the gaps $\Delta_1$ and $\Delta_2$ decrease towards the center in parallel with roughly the same length scale $\xi_1\sim \xi_2 \sim \SI{45}{\nano\metre}$. In the whole range, $\Delta_2$ follows a simple tanh function. At 50 nm, however, $\Delta_1$ crosses $\Delta_2$ and stays larger than $\Delta_2$. Thus, for the band 1, the gap size deviate from the tanh function near the center and decreases on a shorter length scale. This observation for a single band superconductor is known as the Kramer-Pesch effect \cite{kramer_core_1974}. It was quantified by numerical calculations by Gygi and Schlüter \cite{gygi_self-consistent_1991} for the vortex core size of type-II superconductors.
The slope of $\Delta_1$ near the core corresponds to a core size $\xi^{(c)}_1 = \Delta_1(\infty)\left[\lim_{r\rightarrow 0} \frac{\mathrm{d}\Delta_1(r)}{\mathrm{d}r}\right]^{-1} $of only $ \sim \SI{10}{\nano\metre}$. 
In our self-consistent calculation of the pair-potential $\Delta(r)$ for an isotropic vortex in the quasiclassical theory (see Supplementary Material), this Kramer-Pesch shrinking effect is also present and leads to substantial deviation from a tanh function with one universal $\xi$. 
Note that theory in the clean limit, however, predicts a core shrinking proportional to $T/T_c$ when lowering the temperature. At very low temperatures $T\ll T_c$ the slope of the order parameter d$\Delta$/d$r$ at the vortex centre is even predicted to become infinite, which would show as a jump of $\Delta(r)$ at $r=0$ that is smoothed out over the distance $\xi T/T_c$ \cite{volovik_vortex_1993}. 
Experimentally, we do not find this extreme shrinking. 
The "squeezing" of low angular momentum states in the vortex core and thus the Kramer-Pesch effect is absent for the star, i.e. CdGM states of band 2
. Consequently, $\xi^{(c)}_1$ deviates from $\xi^{(c)}_2$ which indicates that intra-band coupling dominates over inter-band coupling, i.e. the two bands are sufficiently decoupled from each other, despite the gap sizes being not too different \cite{ichioka_locking_2017}. For an isotropic two-band superconductor, the shrinking of the core region for only one of the bands has also been predicted in the case of rather weakly coupled bands \cite{silaev_microscopic_2011}. Again, this hints towards a low inter-band coupling in Pb.

\subsection{Anomalous single-flux-quantum vortices}
\begin{figure*}[pt]
    \centering
    \includegraphics[width=\textwidth]{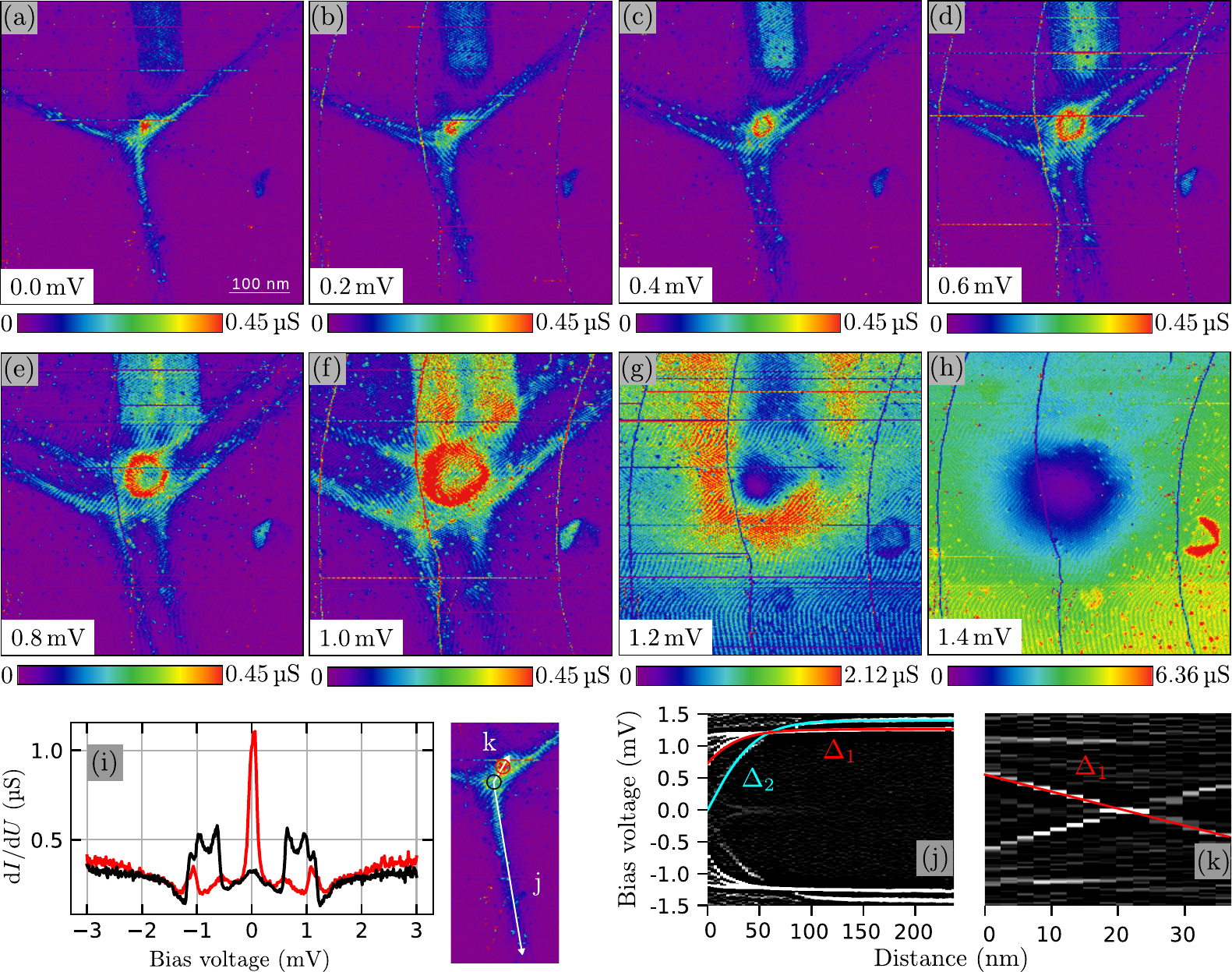}
    \caption{\textbf{Anomalous vortex signature.} (a-h) d$I/$d$U$ maps ($B=\SI{19}{\milli\tesla}$) at different bias voltage for an anomalous vortex (compared to the normal vortex configuration from Fig.~\ref{fig:single_flux_vortex}). (i) Right panel: Enlarged zero bias d$I$/d$U$ map of the anomalous vortex including the position of line spectra (white arrows) and single spectra locations (red/black circle). Left Panel: Single d$I$/d$U$ spectra in the star centre (black) and ring centre (red) revealing ZBPs of different amplitude. (j,k) Heat maps of the second derivative of the differential conductance $-\partial^2 \sigma/\partial U^2$ along the cross-sections marked in (i).
    Red and cyan lines follow $\Delta_1$ and $\Delta_2$ respectively.
    }
    \label{fig:anomalous_vortex}
\end{figure*}
Besides the regular vortex situation, we find vortices, in which the two sets of CdGM states are laterally displaced. Figure \ref{fig:anomalous_vortex} shows such an anomalous vortex.
Both the ring centre and the star centre of the CdGM states are independently movable by a change in magnetic field and their relative displacement can be manipulated into different configurations, even back to the normal configuration from Figure \ref{fig:single_flux_vortex} (see Supplementary Material). We explain this by two effects. First, a change in magnetic field laterally moves the vortex and with it, the two sets of CdGM states. Second, a change in magnetic field can lead to a tilting or bending of the flux lines away from a normal direction to the surface. This leads to a breaking of the cylinder symmetry and can displace the two sets of CdGM states relative to each other. The individual sets of CdGM states behave as those of the regular vortices, except for their relative displacement (see Figure \ref{fig:anomalous_vortex} (a-h)). 

Interestingly, the lateral displacement allows an independent probing of the state sets and thus, an independent identification of their bands. By looking at single bias spectra at the star centre (black) and the ring centre (red) in Figure \ref{fig:anomalous_vortex}(i) we find that the amplitude of the peak at zero bias is three times larger for the ring than for the star, which is supported by our separate band simulations from earlier and can be explained by the larger lateral confinement of low angular momentum states in band 1 compared to band 2. Figures \ref{fig:anomalous_vortex}(j,k) reveal maxima in the LDOS along the cross-sections marked in the right panel of (i) in the form of heat maps of the differential conductance's second derivative with respect to bias voltage. It becomes apparent that $\Delta_2$ (cyan line) closes entirely whereas $\Delta_1$ (red line) does not completely close in the star's centre (black circle). Instead, $\Delta_1$ (red line) closes about $\SI{20}{\nano\metre}$ away from the star centre (red circle). Consequently, the superconducting gap $\Delta_1$ is linked to the ring and $\Delta_2$ to the star. The fact that the quasiparticles of $\Delta_1$ and $\Delta_2$ can be independently displaced with respect to each other rules out rigid inter-band coupling.

\subsection{Multiple-flux-quanta vortices}
\begin{figure*}[pt]
    \centering
    \includegraphics[width=\textwidth]{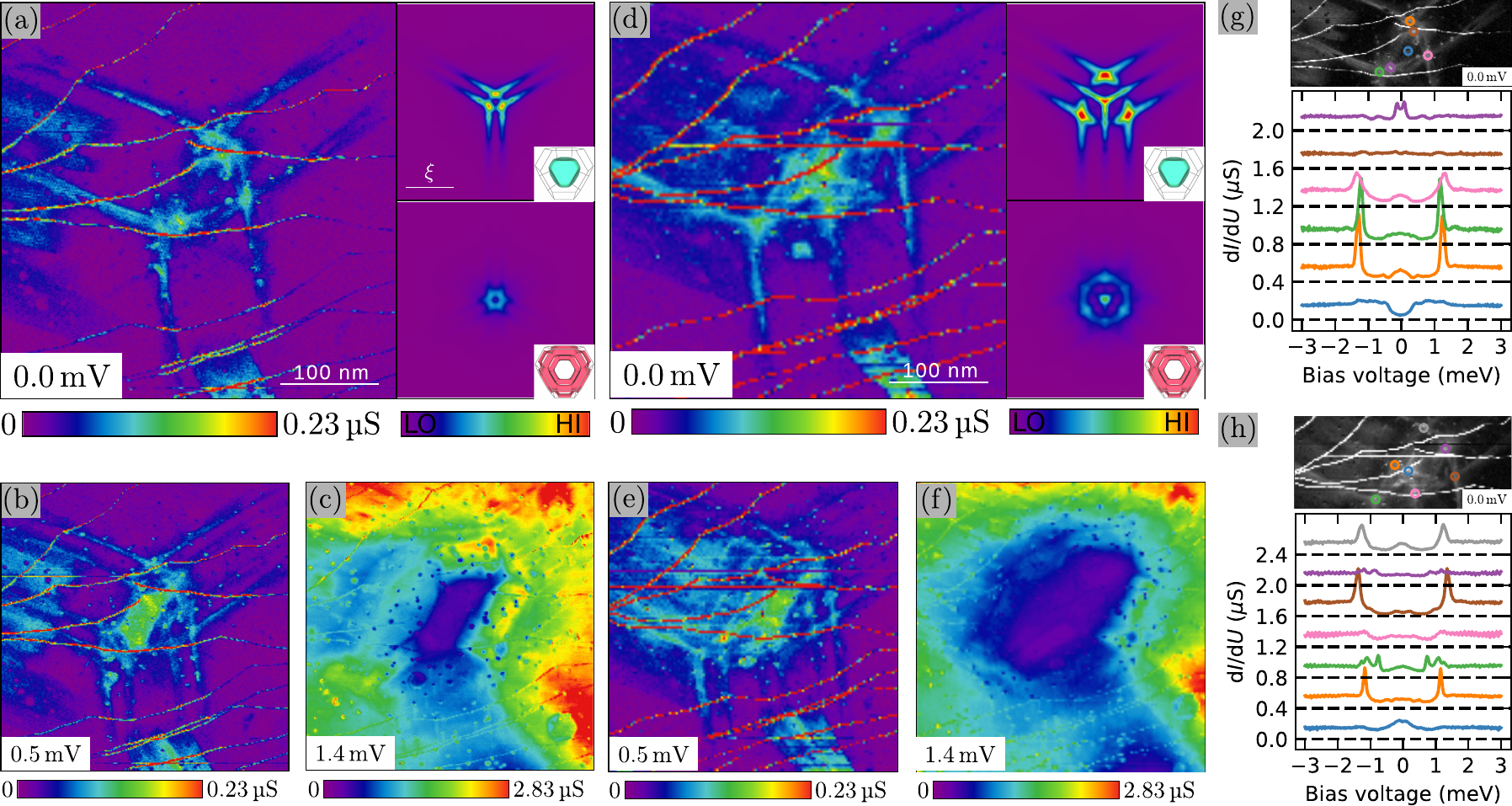}
    \caption{\textbf{Multiple-flux-quanta vortex signature.} (a-f) d$I$/d$U$ maps of an $m=2$ vortex at $B=\SI{0}{\milli\tesla}$ (a-c) and an $m=3$ vortex at $B=\SI{33}{\milli\tesla}$ (d-f) at selected bias voltages. Right panels in (a) and (d) show the simulated LDOS of the bands 1 and 2 for the respective $m$ quanta vortex and energy. (g,h) Bias spectroscopies (bottom panel) recorded with the tip at specific locations marked in the zero bias maps (top panel) of the $m=2$ (g) and $m=3$ (h) vortex. Individual spectra are offset by $\SI{0.4}{\micro S}$ for clarity. Black dashed lines indicate their respective zero conductance baselines.}
    \label{fig:multi_flux_vortex}
\end{figure*}
We also observed larger vortices in the experiments. Figure \ref{fig:multi_flux_vortex} displays two examples. (a-c) shows a vortex with two flux quanta. Our semiclassical calculations indicate that for each flux quantum and each band, a branch of CdGM states is present. As a result, for the vortex with two flux quanta, the band 2 causes a structure with two arms per direction at zero bias (see Fig.~\ref{fig:multi_flux_vortex}(a)) that individually split into two arms with bias voltage (see Fig.~\ref{fig:multi_flux_vortex}(b)) just as in the single-flux-quantum vortex. An analogous behaviour is observed in the vortex with three flux quanta shown in Fig.~\ref{fig:multi_flux_vortex}(d-f). Both vortices are larger than the single flux quantum vortex (compare Fig.~\ref{fig:single_flux_vortex}(h) with Fig.~\ref{fig:multi_flux_vortex}(c,f)). Further, they deviate from a round shape and laterally grow with the number of flux quanta. The number of flux quanta in the vortices does not seem to be limited. We observed giant vortices with over 10 flux quanta (see Supplementary Material). For band 1, the problem of multiple-flux-quanta vortices is similar to that of a single band superconductor with a spherical Fermi surface and has been studied in detail by Volovik {\it et al.} \cite{volovik_localized_1991,volovik_vortex_1993}. In essence, an $m$-flux quanta vortex results in $m$ branches of circular CdGM states of different radii. Due to symmetry, at zero bias and for odd $m$, a central spot is formed by one branch of the CdGM states and the other states form pairs of increasing ring diameters. The CdGM states at energies away from the Fermi level evolve by changing the radii. Thus, the central CdGM state turns from a spot of zero radius to a ring of finite radius, and the pairs of CdGM states split in their radius.
For an even $m$, no central spot is present at the Fermi energy but only pairs of CdGM states with distinct radii exist, that again split when moving away from the Fermi energy. This is the essence of the topological index number theory of Volovik.
In our case, we find a central spot of high differential conductance at zero bias voltage for the vortices with $m=1$ and $m=3$, while there is no central spot for the vortex with $m=2$ in agreement with the topological index number theory. When going away from the Fermi energy, the ring-shaped and spot-like CdGM states overlap with the star shaped states such that their distinction becomes impractical. Tunneling spectra at selected locations inside the vortex showing the very same effects are shown in Figure \ref{fig:multi_flux_vortex}(g,h). 

While the original formulation of Volovik's index number theorem is not directly applicable in real space (that is counting the number of zero modes in the vortex core as function of radius) in our case of an anisotropic Fermi surface, we still believe that the number of parallel arms of our star-shaped CdGM state pattern at zero energy exactly reflects the winding number $m$ of the vortex. The reason for our confidence lies in the fact that the diabolical points in $k$-space, that need to be counted, still lie at zero energy, it is just not obvious which path in real space crosses each of them exactly once. Comparing the axisymmetric problem with the problem at hand, it becomes obvious that a parametrization in terms of the set of quantum numbers $(k_r, k_\phi, k_z)$ has to be replaced by an irreducible representation of the crystallographic point group. The flat parts of Fermi surface 2 focus the quasiparticles into high-symmetry directions and instead of a localization of different CdGM state branches at different radii, they now localize at different impact parameters. 

At last, we tested whether vortices would also be present at temperatures much closer to $T_c$. Our measurements at $\SI{4.3}{\kelvin}$ revealed that it is substantially harder to trap a vortex in our scan frame, but we managed to in a single case (see Supplementary Material).

\section{Conclusions}
In conclusion, we report the observation of single-flux-quantum and multiple-flux-quanta vortices in a traditional type-I bulk superconductor, i.e. single crystal Pb(111), by low-temperature scanning tunneling microscopy. We demonstrate a robust determination method for the winding number of the vortex by usage of a topological index theorem which relates the number of flux quanta and CdGM state branches.
The spatial anisotropy of the quasiparticle states inside the vortex reflects the crystal symmetry, and its shape is governed by the anisotropic Fermi velocity in the superconducting bands. An influence of neighboring flux lines can be ruled out due to the absence of an ordered flux pattern. In addition, we could show how CdGM states from two weakly coupled bands interact with each other in a single flux line.
While the generically unavoidable inter-band coupling leads to the formation of a single coherence length and order parameter right at $T_c$ \cite{kogan_ginzburg-landau_2011}, the physics below $T_c$ is considerably richer, in particular in the regime of weak inter-band coupling \cite{silaev_microscopic_2011, ichioka_locking_2017}. Of particular interest is the emergence of mixed gap modes \cite{silaev_microscopic_2011} or the possibility of the observation of the rather elusive Leggett mode in the limit of weak inter-band coupling \cite{leggett_number-phase_1966}, i.e. a collective fluctuation of the relative phase of the two bands, for superconducting Pb. We experimentally demonstrate that for systems with multi-band pairing, the transitional region of the hysteretic behaviour expands due to microscopic interactions and also allows for interesting vortex configurations, like vortex clusters or multiple-flux-quanta vortices \cite{dao_giant_2011}. 
The emergence of such vortices in the intermediate state of a prototypical type-I superconductor is very surprising and 
demonstrates that the low temperature flux patterns of two-band superconductors rank on a spectrum between type I and II. Not only does the existence of vortices in Pb allow for an alternative test ground of vortex physics in multi-component superconducting systems apart from MgB$_2$, the variability of the vortex winding number also suggests a combination with topological crystal defects that was recently predicted to result in topological quasiparticle states, like Majorana zero modes, under certain circumstances \cite{samokhvalov_topological_2020, rex_topological_2022}.

\section*{Methods}
\paragraph{Experimental details}:
The experiments were performed with a home-built scanning tunneling microscope with dilution refrigeration, which can reach a base temperature of $\SI{25}{\milli\kelvin}$ in a magnetic field of up to $\SI{7.5}{\tesla}$ \cite{Balashov2018}. 
In our setup, the bias voltage $U_t$ is applied between sample and common machine ground so that a positive bias voltage probes the unoccupied states of the sample. The STM chamber is kept at a base pressure of $\SI{1e-10}{\milli\bar}$. The single crystal Pb(111) (miscut angle: $\pm 0.1^\circ$,  purity: $99.999\%$) has been purchased from MaTecK GmbH. It has cylindrical (hat) shape with a diameter of $\SI{8}{\milli\metre}$ and a thickness of $\SI{2}{\milli\metre}$. At a base pressure of $\SI{1e-10}{\milli\bar}$ the Pb crystal was prepared in cycles of sputtering with Ar$^+$ ions of $\SI{3}{\kilo\electronvolt}$ and subsequent annealing at $190^{\circ}$C and directly transferred into the STM in-situ. A tungsten tip was prepared by high-temperature flashing and soft dipping into a Au(111) surface in order to avoid picking up Pb atoms. The measurements (except for the one at $\SI{4.3}{\kelvin}$
) were all performed below $\SI{45}{\milli\kelvin}$. After zero-field cooling the Pb crystal, the vortices were formed by ramping the perpendicular magnetic field from $\SI{0}{\milli\tesla}$ to $\SI{85}{\milli\tesla}$ and back down to a constant value below the critical field. Note that $\SI{80}{\milli \tesla}$ is the critical magnetic field for Pb \cite{chanin_critical-field_1972}. 
The differential conductance was measured using a Lock-in amplifier at a frequency of $3.4-3.6\,\SI{}{kHz}$ and AC peak amplitude $U_\mathrm{ac}^\mathrm{PK}$ between $10$ and $\SI{100}{\micro\volt}$ (for details, see Supplementary Material). d$I$/d$U$ maps at sub-gap energies were recorded in a multi-pass configuration: In the ``record" phase the tip records the $z$-profile at the feedback condition (constant tunneling current of $I_t$ at $eU_t>\Delta$) and in the ``play" phase the $z$-profile is repeated at a different bias voltage. In order to increase the signal for sub-gap energies an offset is often added to the $z$-profile bringing the tip closer to the surface. This record and play phase alternation is performed line for line until the entire area has been scanned.

\paragraph{Calculation details}:
In order to obtain the simulated LDOS for vortices containing an arbitrary number of flux quanta, we used the Riccati parametrisation of the quasiclassical 3D Eilenberger equations as proposed in Ref. \cite{schopohl_transformation_1998} and numerically solved the one-dimensional differential equations under appropriate boundary conditions (for details, see Supplementary Material) for each band separately. Motivated by our experimental findings, we assumed a radial symmetric local pair potential $\Delta(r)$ in the plane perpendicular to a vortex line with s-wave symmetry that vanishes in the vortex centre. We used the ratio $\Delta_2(\infty)/\Delta_1(\infty)$ obtained from the experiment. We respected the broken translation symmetry at the crystal surface by a work function term. The magnetic vector potential was set to zero for all calculations shown in the main text. An inclusion of a magnetic vector potential of appropriate form only yielded small quantitative deviations from the zero-field case (see Supplementary Material), yet drastically increased the required computation time, which is why we refrained from it for the LDOS maps. 

Density functional calculations of the electronic structure of Pb
were carried out in the framework of the mixed-basis pseudopotential
method \cite{meyer97}. The electron-ion interaction was
represented by norm-conserving relativistic pseudopotentials \cite{vande85}. Spin-orbit coupling was incorporated within the pseudopotential scheme via Kleinman's formulation \cite{kleinman80} and was consistently taken into account in the charge self-consistency cycle using a spinor representation of the wave functions. Further details of the spin-orbit coupling implementation within the mixed-basis pseudopotential method can be found in a previous publication \cite{heid10}. For higher accuracy, 5$d$ semicore states were included in the valence space.  The deep $d$ potential is efficiently treated by the mixed-basis approach, where valence states are
expanded in a combination of plane waves and local functions.  Here,
local functions of $d$ type at the Pb sites were combined with plane
waves up to a kinetic energy of 20~Ry.  Brillouin-zone integration
was performed by sampling a 32$\times$32$\times$32 $k$-point mesh (corresponding to 2992 $k$ points in the irreducible part of the Brillouin zone) in conjunction with a Gaussian broadening of 0.2~eV. The
exchange-correlation functional was represented by the local-density
approximation in the parameterization of Perdew-Wang \cite{perdew92}.

This DFT technique was applied to obtain Fermi surface properties entering the Eilenberger equations.
Band energies were calculated on fine radial grids for a cylindrical coordinate system taking the [111] direction as the z-axis, to determine Fermi momenta $k_F$ for each of the two relevant bands.
At each $k_F$, 3-dimensional Fermi velocities $v_F$ were then calculated taking numerical derivatives of band energies around this point.
The optimized lattice parameter $a=4.89$~\AA\, was used throughout.

\section*{Author contributions}
T.G. and Q.L. conducted the experiments, performed the quasiclassical (Eilenberger) calculations and analysed the data. T.G. was responsible for presentation of the data. R.H. performed the DFT calculations. R.N. and T.K.Y. performed initial experiments and provided the crystal. R.W. and J.S. gave theoretical support. W.W. headed the study. T.G. and W.W. wrote the manuscript including input from all authors.
\section*{Acknowledgments}
We acknowledge funding from the Deutsche Forschungsgemeinschaft (DFG) with grant Wu 394/12-1 and discussions with A. Ustinov and S. Rex. R.H. acknowledges support by the state of Baden-Württemberg through bwHPC. This work was supported by JSPS KAKENHI under the grant number 17K19023.

\bibliography{main}

\clearpage


\end{document}


\title{Supplementary Material to ``Identification of multiple-flux-quanta vortices by core states in the two-band superconductor Pb''}

\author{Thomas Gozlinski}
\email{thomas.gozlinski@kit.edu}
\thanks{These authors contributed equally to this work.}
\affiliation{Physikalisches Institut, Karlsruhe Institute of Technology,
Wolfgang-Gaede-Str.1, 76131 Karlsruhe, Germany}
\author{Qili Li}
\thanks{These authors contributed equally to this work.}
\affiliation{Physikalisches Institut, Karlsruhe Institute of Technology,
Wolfgang-Gaede-Str.1, 76131 Karlsruhe, Germany}
\author{Rolf Heid}
\affiliation{Institute for Quantum Materials and Technologies, Karlsruhe Institute of Technology, Hermann-von-Helmholtz-Platz 1, 76344 Eggenstein-Leopoldshafen, Germany}
\author{Ryohei Nemoto}
\affiliation{Department of Materials Science, Chiba University,
1-33 Yayoi-cho, Inage-ku, Chiba 263-8522, Japan}
\author{Roland Willa}
\affiliation{Institute for Theory of Condensed Matter, Karlsruhe Institute of Technology,
Wolfgang-Gaede-Str.1, 76131 Karlsruhe, Germany}
\author{Toyo Kazu Yamada}
\affiliation{Department of Materials Science, Chiba University,
1-33 Yayoi-cho, Inage-ku, Chiba 263-8522, Japan}
\affiliation{Molecular Chirality Research Centre, Chiba University,1-33 Yayoi-cho, Inage-ku, Chiba 263-8522, Japan}
\author{Jörg Schmalian}
\affiliation{Institute for Quantum Materials and Technologies, Karlsruhe Institute of Technology, Hermann-von-Helmholtz-Platz 1, 76344 Eggenstein-Leopoldshafen, Germany}
\affiliation{Institute for Theory of Condensed Matter, Karlsruhe Institute of Technology,
Wolfgang-Gaede-Str.1, 76131 Karlsruhe, Germany}
\author{Wulf Wulfhekel}
\affiliation{Physikalisches Institut, Karlsruhe Institute of Technology,
Wolfgang-Gaede-Str.1, 76131 Karlsruhe, Germany}
\affiliation{Institute for Quantum Materials and Technologies, Karlsruhe Institute of Technology, Hermann-von-Helmholtz-Platz 1, 76344 Eggenstein-Leopoldshafen, Germany}


\date{\today}


\maketitle 

\clearpage

\renewcommand{\thefigure}{S\arabic{figure}}
\setcounter{figure}{0} 
\renewcommand{\thetable}{S\arabic{table}}
\setcounter{table}{0} 


\section*{Contents}
\noindent Extended Methods\\
Supplementary Notes 1-5\\

\section*{Extended methods}

\paragraph{STM measurement conditions}
Tab. \ref{tab:stm_conditions} summarizes the most important STM/STS measurement parameters used to obtain the figures in the main text and supplementary material.

\begin{table}[hpb]
    \centering
    \caption{\textbf{Measurement parameters.} $I_t$ denotes the feedback condition of the tunnelling current at the bias voltage $U_t$. $U_\mathrm{ac}^\mathrm{PK}$ is the peak AC voltage amplitude added via the Lock-in amplifier. In multi-pass d$I$/d$U$ maps $z_\mathrm{off}$ is the distance the tip is brought closer to the surface compared to the feedback condition. $B$ is the magnetic flux density in vacuum and $T$ is the temperature.}
    \setlength\extrarowheight{-3pt}
    \begin{tabular}{l c| c c c c c c}
        \toprule
         Fig.& & $I_t(\SI{}{\nano\ampere})$ & $U_t(\SI{}{\milli\volt})$ & $U_\mathrm{ac}^\mathrm{PK}(\SI{}{\micro\volt})$ &  $z_\mathrm{off}(\SI{}{\pico\metre})$ & $B(\SI{}{\milli\tesla})$ & $T(\SI{}{\milli\kelvin})$\\ \hline
         1& A & $0.2$ & $1$ & - & - & - & $<45$\\
         & B & $0.05$ & $1.3$ & $50$ & - & $23$ & $<45$\\
         & C & $1$ & $3$ & $20$ & - & $0$ & $39$\\
         & D & $2$ & $3$ & $20$ & - & $23$ & $48$\\\hline
         2& A-G & $0.5$ & $1.4$ & $50$ & $20$ & $0$ & $<45$\\
         & H & $0.5$ & $1.4$ & $50$ & - & $0$ & $<45$\\\hline
         3& & $0.5$ & $3$ & $20$ & - & $0$ & $<45$\\\hline
         4& A-G & $1$ & $1.4$ & $50$ & $20$ & $19$ & $<45$\\
         & H & $1$ & $1.4$ & $50$ & - & $19$ & $<45$\\
         & I & $1$ & $3$ & $20$ & - & $19$ & $40$\\
         & J & $1$ & $1.5$ & $10$ & - & $19$ & $40$\\
         & K & $1$ & $2$ & $20$ & - & $19$ & $40$\\\hline
         5& A-B & $0.5$ & $1.4$ & $50$ & $20$ & $0$ & $<45$\\
         & C & $0.5$ & $1.4$ & $50$ & - & $0$ & $<45$\\
         & D-E & $1$ & $1.4$ & $50$ & $20$ & $33$ & $<45$\\
         & F & $1$ & $1.4$ & $50$ & - & $33$ & $<45$\\
         & G & $1$ & $3$ & $20$ & - & $0$ & $<45$\\
         & H & $1$ & $3$ & $20$ & - & $33$ & $<45$\\\hline
         \ref{fig:anomolous_b_field_dep}& A-D & $1$ & $1.4$ & $50$ & - & IND$^a$& $<45$\\\hline
         \ref{fig:anomolous_to_normal}& A-B & $0.2$ & $1.4$ & $50$ & - & IND & $<45$\\
         & C-D & $1$ & $1.4$ & $20$ & - & IND & $<45$\\\hline
         \ref{fig:giant_flux_vortex}& A-B & $1$ & $3$ & $50$ & $20$ & $0$ & $<45$\\\hline
         \ref{fig:vortex_neg_Bfield_or_bias}& A-B & $0.5$ & $1.4$ & $50$ & - & IND & $36$\\
         & C-D & $1$ & $1.4$ & $20$ & - & IND & $42$\\\hline
         \ref{fig:4K_vortex}& A & $0.1$ & $1.8$ & $50$ & - & $0$ & $4300$\\
         & B & $0.1$ & $1.8$ & $50$ & $20$ & $0$ & $4300$\\
         & C & $0.1$ & $6$ & $100$ & - & $0$ & $4300$\\
    \end{tabular}
    \label{tab:stm_conditions}
    \\\footnotesize{$^a$ This parameter is indicated in the sub-figure itself.} 
\end{table}

\paragraph{3D Eilenberger calculations}
We follow Ref. \cite{schopohl_transformation_1998} in notation and describe how we solved the quasiclassical Eilenberger equations \cite{eilenberger_transformation_1968, larkin_quasiclassical_1969}
\begin{align}
    -\hbar \bm{v}_F \nabla \hat{g}(\bm{r};\bm{p}_F, i \epsilon_n) = 
    \left[
    \begin{pmatrix}
        i \epsilon_n + \bm{v}_F e\bm{A}(\bm{r}) & -\Delta (\bm{r},\bm{p}_F) \\
        \Delta^\dagger (\bm{r},\bm{p}_F) & -i \epsilon_n - \bm{v}_F e\bm{A}(\bm{r})
    \end{pmatrix},
     \hat{g}(\bm{r};\bm{p}_F, i \epsilon_n)
    \right],
    \label{eq:3DEilenberger/Eilenberger transport eq}
\end{align}
that hold for $k_F\xi\gg 1$.
The quasiclassical Green's function propagator
\begin{align}
    \hat{g} = \begin{pmatrix}
       g(\bm{r},\bm{p}_F,i \epsilon_n) & f(\bm{r},\bm{p}_F,i \epsilon_n)\\
       -f^\ast(\bm{r},-\bm{p}_F,i \epsilon_n) & g^\ast(\bm{r},-\bm{p}_F,i \epsilon_n)
    \end{pmatrix},
\end{align}
depends on spatial coordinate $\bm{r}$, crystal momentum $\bm{p}_F=\hbar \bm{k}_F$ and energy $\epsilon_n$, and must satisfy the normalization condition $\hat{g}^2 = \hat{1}$. $g$ and $f$ are normal and anomalous quasiclassical Green's function propagators and $\epsilon_n$ are fermionic Matsubara frequencies. Incorporated in this formalism is the self-consistent calculation of pair potential $\Delta(\bm{r})$ and $\bm{A}(\bm{r})$, which is essential for the description of a superconductor hosting vortices as these two fields are dependent on each other. Although this approach is numerically much more feasible than direct diagonalization of a BdG Hamiltonian when treating inhomegeneties, it is not necessary to solve the problem completely self-consistently in this work. The reason for that is, that our experimental results clearly show, that even though the sub-gap states in the vortex display highly anisotropic behaviour, the recovery of $\Delta(\bm{r})$ is isotropic in the plane around the vortex core. The local pair potential is assumed to have s-wave symmetry and was therefore modeled by 
\begin{align}
    \Delta(\bm{r},\bm{p}_F) = \Delta(\bm{r}) = \left(\Delta_0\tanh \frac{r}{\xi} + \Theta(z) W\tanh \frac{z}{a}\right)\left(\frac{x+i y}{r}\right)^m, 
\end{align}
with $\Theta$ being a Heaviside step function, $\Delta_0$ the maximum gap size, $\xi$ the coherence length, $a$ the lattice constant, $m$ the winding number of the vortex and $W$ the work function. The second term ensures, that quasiparticles travelling to the surface are decaying into the vacuum with the right damping factor. Since an isotropic $\Delta$ implies an isotropic in-plane current density, the magnetic field profile around the vortex was described by a vector potential of the form \cite{carneiro_vortex_2000}
\begin{align}
    \bm{A} &= A(r,z) \hat{\bm{e}}_\varphi \,,\\
    A(r,z) &= \frac{m\Phi_0}{2\pi\lambda^2}\int\limits_0^\infty \mathrm{d}k \frac{J_1(kr)}{k^2+\lambda^{-2}}S(k,z) \,,\label{eq:app2/Vector_potential}\\
    S(k,z) &= 
    \begin{cases}
        \frac{\kappa}{k+\kappa}\mathrm{e}^{-kz} & z>0\\
        1-\frac{k}{k+\kappa}\mathrm{e}^{\kappa z} & z\leq 0 \,,
    \end{cases}
\end{align}
that is cylinder symmetric in the bulk and deviates from the bulk value near and above the surface. Here, $\kappa = \sqrt{k^2+\lambda^{-2}}$, $\lambda$ is the magnetic penetration depth which was chosen to be $\lambda=\xi/\sqrt{2}$ and $J_1(x)$ is a Bessel function of first order.

Nils Schopohl and Kazumi Maki \cite{schopohl_quasiparticle_1995} could show that the Eilenberger equations can always be solved along a characteristic line and that the solution is universal for each point along this line. The line simply has to be parallel to the Fermi velocity vector $\bm{v}_F$. Points on this line are then characterized by the variable $X$ and two \textit{impact parameters} $Y$ and $Z$. 
\begin{align}
    \bm{r}(X) &= X \hat{\bm{u}} + Y \hat{\bm{v}} + Z \hat{\bm{w}},\\
    &= x \hat{\bm{x}} + y \hat{\bm{y}} + z \hat{\bm{z}} 
\end{align}
Transformation from the mobile frame of reference $(\hat{\bm{u}},\hat{\bm{v}},\hat{\bm{w}})$, where $\hat{\bm{u}}\parallel \bm{v}_F$, to the fixed coordinate system $(\hat{\bm{x}},\hat{\bm{y}},\hat{\bm{z}})$ is done by chaining rotation matrices using Euler angles:
\begin{align}
    \begin{pmatrix}
        v_x\\ v_y\\ v_z
    \end{pmatrix}&=
    R_{zyz}
    \begin{pmatrix}
        v_X\\ v_Y\\v_Z
    \end{pmatrix}\\
    R_{zyz} &= R_z(\eta) R_y(\chi - \pi/2) R_z(\psi) \nonumber\\
    &= 
    \begin{pmatrix}
        \cos\eta & -\sin\eta & 0\\
        \sin\eta & \cos\eta & 0 \\
        0 & 0 & 1
    \end{pmatrix}
    \begin{pmatrix}
        \sin\chi & 0 & \cos\chi\\
        0 & 1 & 0\\
        -\cos\chi & 0 & \sin\chi 
    \end{pmatrix}
     \begin{pmatrix}
        \cos\psi & -\sin\psi & 0\\
        \sin\psi & \cos\psi & 0 \\
        0 & 0 & 1
    \end{pmatrix}.
\end{align}
In order to align the axis $\hat{\bm{u}}$ of the mobile frame with the axis $\hat{\bm{x}}$ of the static coordinate system, no rotation about $\hat{\bm{w}}$ is needed and thus $\psi=0$. The other two rotational angles are defined by the components of the Fermi velocity vector as follows:
\begin{align}
    \eta &= \arccos \frac{v_z}{|\bm{v}|} \,,\\
    \chi &=
    \begin{cases}
        \arctan \frac{v_y}{v_x}\,, & v_x>0\\
        \arctan \frac{v_y}{v_x}+\pi\,, & v_x<0\\
        \text{sgn}(v_y)\frac{\pi}{2}\,, & v_x=0
    \end{cases}.
\end{align}
Using this transformation, the vector potential and gap parameter can be expressed in the variables of the mobile frame: $A(\bm{r}(X,Y,Z))$ and $\Delta(\bm{r}(X,Y,Z))$. On a characteristic line defined by the impact parameters $Y_p$ and $Z_p$, the functions from Eq.~\eqref{eq:3DEilenberger/Eilenberger transport eq} become \cite{schopohl_transformation_1998}
\begin{align}
    \Delta(X) &= \Delta(\bm{r}(X,Y_p,Z_p)) \,,\\
    i \Tilde{\epsilon}_n(X) &= i \epsilon_n + \bm{v}_F e\bm{A}(\bm{r}(X,Y_p,Z_p)) \,,
    \label{eq:app2/New_Matsubara_frequencies}\\
    \hat{g}(X) &= \hat{g}(\bm{r}(X,Y_p,Z_p),\bm{p}_F,i \epsilon_n).
\end{align}
The parametrisation of the Eilenberger equations on this 1D line is called the \textit{Riccati parametrisation}. Eq.~\eqref{eq:3DEilenberger/Eilenberger transport eq} formally reduces to the solution of two initial value problems, where the differential equations are scalar and of first order. The Eilenberger propagator is parametrised in terms of two scalar complex functions $a(X)$ and $b(X)$:
\begin{align}
    \hat{g}(X) = 
    \frac{1}{1+a(X)b(X)}
    \begin{pmatrix}
       1-a(X)b(X) & 2 i a(X)\\
       -2i b(X) & -1+a(X)b(X)
    \end{pmatrix}.
\end{align}
The differential equations that need to be numerically solved for $a(X)$ and $b(X)$ are \cite{schopohl_quasiparticle_1995}
\begin{align}
    \hbar v_F a^\prime(X)+[2\Tilde{\epsilon}_n+\Delta^\dagger(X)a(X)]a(X)-\Delta(X)&=0 \,,\\
    \hbar v_F b^\prime(X)-[2\Tilde{\epsilon}_n+\Delta(X)b(X)]b(X)+\Delta^\dagger(X)&=0
\end{align}
with boundary conditions
\begin{align}
    a(-\infty)&=\frac{\Delta(-\infty)}{\epsilon_n+\sqrt{\epsilon^2_n+|\Delta(-\infty)|^2}} \,,\\
    b(+\infty)&=\frac{\Delta^\dagger(+\infty)}{\epsilon_n+\sqrt{\epsilon^2_n+|\Delta(+\infty)|^2}}.
\end{align}
In order to solve them at a certain energy $E<\Delta_0$, the analytical continuation $i\epsilon_n \rightarrow E + i 0^+$ was used.
Finally, the local density of states at $\bm{r}$ was obtained from the real part of $g(\bm{r}(X))$, i.e.
\begin{equation}
    \mathcal{N}(\bm{p}_F) = \mathcal{N}_0(\bm{p}_F)\mathrm{Re}\left(\frac{1-a(X)b(X)}{1+a(X)b(X)}\right).
\end{equation}
In order to obtain the total local density of states one still needs an integration over the Fermi surface to calculate all trajectories that are present due to the various Fermi velocity vectors that exist, plus an integration over the impact factors $Y$ in order to account for trajectories that do not traverse the vortex centre. An integration over $Z$ is redundant since the solution in the bulk is the same for every $Z$. Near the surface this is strictly not the case anymore but the effect is small and with a large enough variety of Fermi velocity vectors (which we have) these trajectories are not missed. 

\begin{figure}
    \centering
    \includegraphics[scale=1]{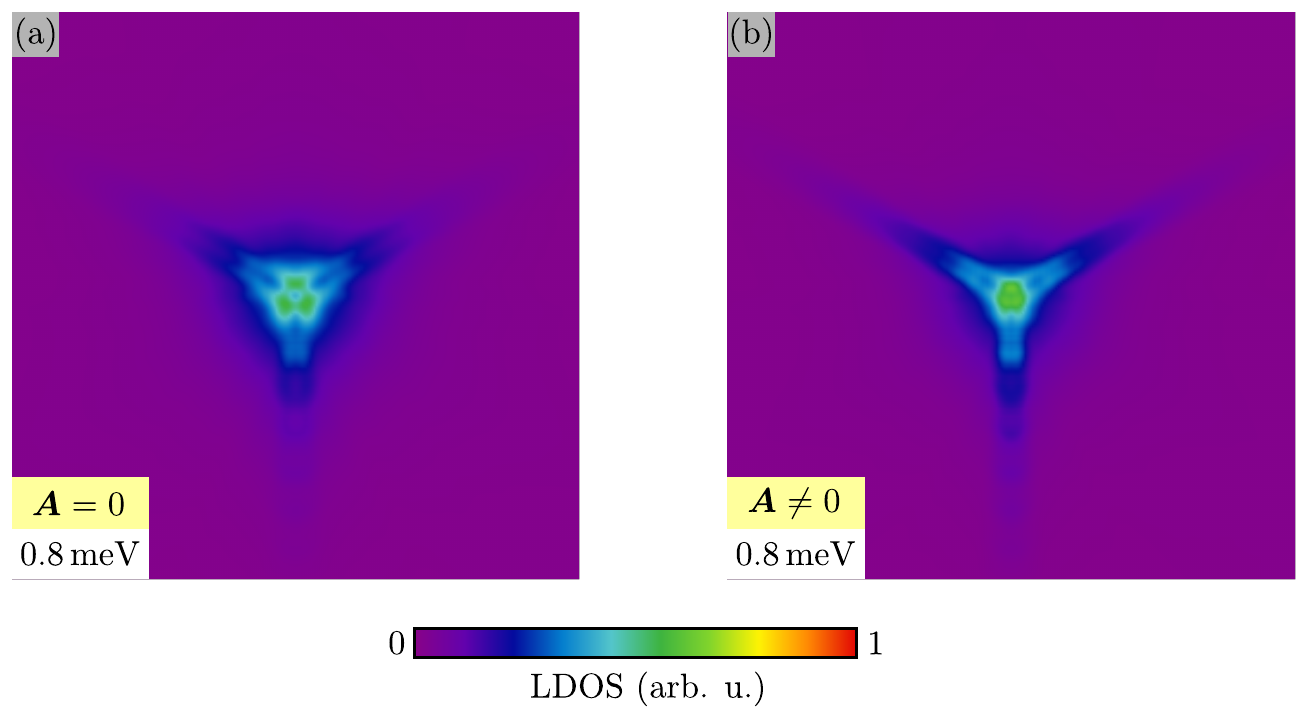}
    \caption[Influence of vector potential]{\textbf{Influence of Vector Potential}: The LDOS obtained from solutions of the 3D Eilenberger equations for a single-flux-quantum vortex at energy $eU=\SI{0.8}{\milli e\volt}$ without (a) and with vector potential (b), as formulated in Eq.~\eqref{eq:app2/Vector_potential}, show only quantitative differences. With non-zero vector field, the star arms are still split, yet the CdGM states at this energy are squeezed into a smaller area around the vortex core.}
    \label{fig:app2/Vector Potential Influence}
\end{figure}

\paragraph{Influence of vector potential}
The inclusion of a non-zero vector potential $\bm{A}$ in the calculations increases the average time needed to solve the Eilenberger equations because the Matsubara frequencies in Eq.~\eqref{eq:app2/New_Matsubara_frequencies} gain a position dependent term that requires a coordinate transformation between the two reference frames. Therefore, the simulations shown in the main text are performed without vector potential. It is already visible from Eq.~\eqref{eq:app2/New_Matsubara_frequencies} that in a calculation where $\bm{A}$ and $\Delta$ are not solved self-consistently, $\bm{A}$ only enters the equation as an effective energy term. With a vector potential in the azimuthal direction like in Eq.~\eqref{eq:app2/Vector_potential} the scalar product with the Fermi velocity is only expected to yield significant contribution for large impact parameters (for $Y_p=0$, $\bm{v}_F$ is perpendicular to $\bm{A}$). That means trajectories with increasing impact parameter have large LDOS already for smaller distances than in the field free case. The splitting star arms in the LDOS maps should be squeezed to smaller distances from the core. In fact, this is what is seen in the calculations with vector potential at higher energies, as shown in Fig.~\ref{fig:app2/Vector Potential Influence}. This proves that even though there are quantitative differences to the case without vector potential, in the general characteristics, the LDOS patterns remain unchanged.

\clearpage

\section*{Supplementary Note 1: Anomalous vortices in varying magnetic fields}
\begin{figure}[hpb]
    \centering
    \includegraphics{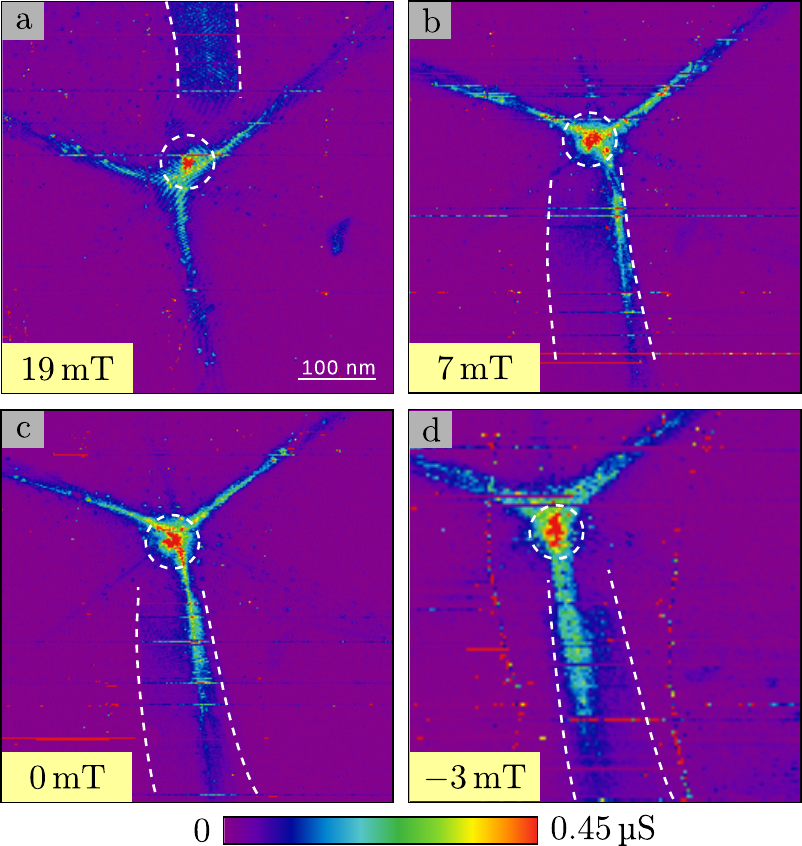}
    \caption{\textbf{Manipulation of anomalous vortex pattern.} Zero bias d$I$/d$U$ maps of an anomalous vortex at different magnetic fields. The relative positional shift between the star centre and ring centre (red) is manifold and not tied to crystal directions. The position of the ring centre is marked by a white-dashed circle. The bright stripe mentioned in the main text is marked at its borders by white dashed lines.}
    \label{fig:anomolous_b_field_dep}
\end{figure}

The direction and magnitude of spatial displacement between the sets of CdGM state branches from the two superconducting bands found in anomalous vortices could not be tied to any crystal direction and appears to be almost random. By slowly varying the static magnetic field, the displacement can be modified, as shown in the d$I$/d$U$ maps of the same anomalous vortex in Fig.~\ref{fig:anomolous_b_field_dep} at different fields. We do, however, see an additional blue stripe in the zero-bias d$I$/d$U$ maps of anomalous vortices that appears on the same side of the star pattern as the ring state. For clarity, the ring state is enclosed by a white-dashed circle and the borders of the mentioned stipe are highlighted by white-dashed contour lines in Fig.~\ref{fig:anomolous_b_field_dep}. These stripes might hint at the direction in which the magnetic field lines are tilted beneath the surface.

\begin{figure}[hpt]
    \centering
    \includegraphics{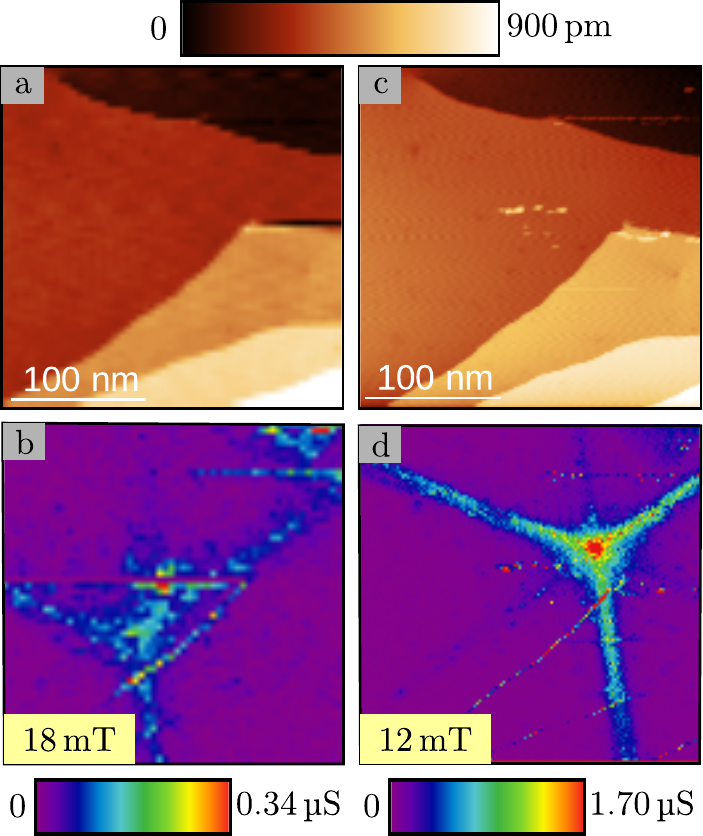}
    \caption{\textbf{Manipulation from anomalous to normal vortex.} (a, c) Topographic images of the corresponding quasiparticle patterns in (b,d) showing the same location on the surface. (b,d) Zero bias d$I$/d$U$ map of the anomalous vortex before (b) and after the magnetic field ramp (d). After the decrease of magnetic field, the anomalous vortex in (b) moved and transformed into a normal type (d). In the process, both sets of CdGM states moved.}
    \label{fig:anomolous_to_normal}
\end{figure}

As Fig.~\ref{fig:anomolous_to_normal} demonstrates, a vortex core can be moved by a slow change in magnetic field and the sets of CdGM states from the two superconducting bands are displaced independently. Consequently, we are able to transform an anomalous vortex back to a normal vortex proving that the two types are not inherently different and that the anomalous type may well be explained by flux line tilting.

\clearpage

\clearpage

\section*{Supplementary Note 2: Giant vortex}
\begin{figure}[hpb]
    \centering
    \includegraphics[width=0.5\textwidth]{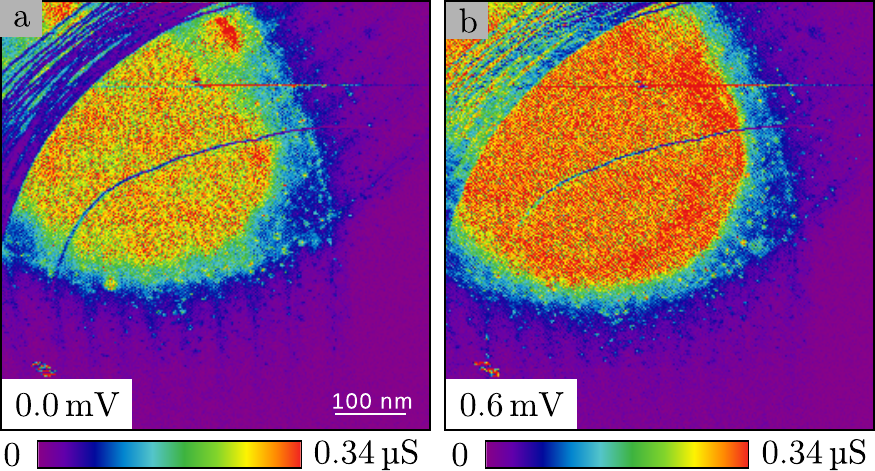}
    \caption{\textbf{Giant vortex.} d$I$/d$U$ maps of a giant vortex containing $m>10$ flux quanta. It shows more than 10 arms at zero bias (a) that individually split in two at sub-gap energies away from the Fermi level (b). This vortex is located at the edge of a large sputtering defect.}
    \label{fig:giant_flux_vortex}
\end{figure}

In a single case, a giant vortex containing $m>10$ flux quanta was found near a large sputtering defect. The zero-bias d$I$/d$U$ map of this vortex, shown in Fig.~\ref{fig:giant_flux_vortex}(a), features more than 10 arms of CdGM states stretching in each $\langle 2\bar{1}\bar{1} \rangle$ direction, which again split into pairs at larger bias (b). Unfortunately this particular giant vortex was partly outside the scan frame of the fine motion piezo tube, which prevented us from a definite determination of $m$.

\clearpage

\section*{Supplementary Note 3: Particle-hole and time-reversal symmetry}
\begin{figure}[hpb]
    \centering
    \includegraphics[width=\textwidth]{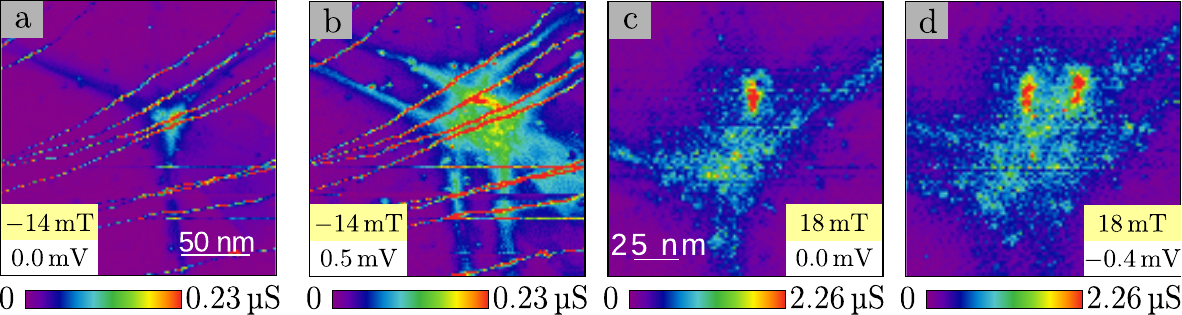}
    \caption{\textbf{Reversal of magnetic and electric field.} Applying the magnetic field or electric field in the opposite direction has no effect on the LDOS pattern inside the vortex. (a-b) d$I$/d$U$ maps of a normal single-flux-quantum vortex in reversed magnetic field. (c-d) d$I$/d$U$ maps of an anomalous vortex showing that reversing the sample bias yields indentical LDOS patterns.}    \label{fig:vortex_neg_Bfield_or_bias}
\end{figure}

We find that the differential conductance maps show no qualitative change (compared to the figures in the main text) when the electric field in the tunnelling junction is reversed (inverse bias voltage sign $U\leftrightarrow -U$) or the magnetic field direction is reversed ($B\hat{\bm{e}}_z \leftrightarrow -B\hat{\bm{e}}_z$). As a point of proof, Fig.~\ref{fig:vortex_neg_Bfield_or_bias}(a,b) shows a normal vortex stabilized at $\SI{-14}{\milli\tesla}$ (after saturation at $\SI{-85}{\milli\tesla}$) for $U\geq \SI{0}{\milli\volt}$ and (c,d) shows an anomalous vortex stabilized at $\SI{18}{\milli\tesla}$ for $U\leq \SI{0}{\milli\volt}$. The invariance of the LDOS maps under reversal of the electric field demonstrates the particle-hole symmetry of the CdGM states and justifies their treatment as excitations of a BCS ground state, i.e. describing their dynamics by a mean-field BdG Hamiltonian or Eilenberger's quasiclassical Green's function propagators. The invariance of the LDOS maps under reversal of the magnetic field is the logical consequence of time-reversal symmetry of the CdGM states.

\clearpage

\section*{Supplementary Note 4: Vortex at $\SI{4.3}{\kelvin}$}
\begin{figure}[hpb]
    \centering
    \includegraphics[width=0.75\textwidth]{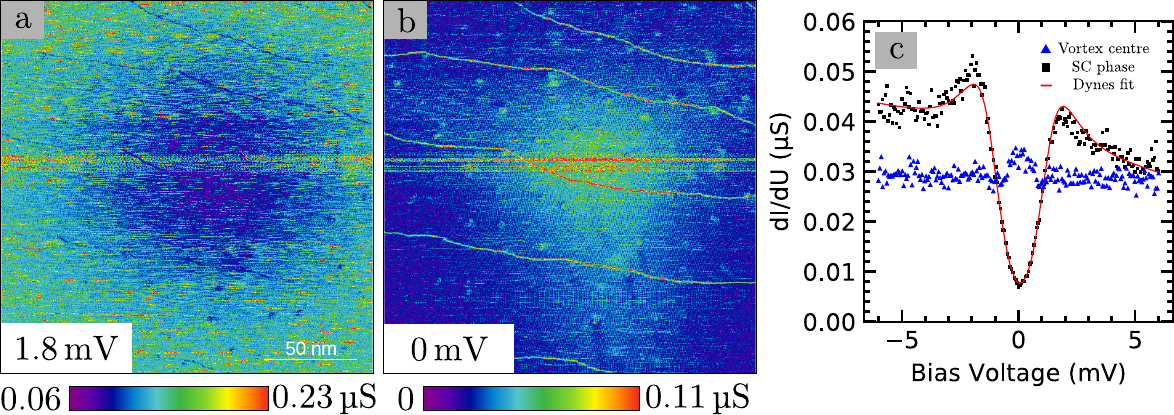}
    \caption{\textbf{Vortex at higher temperature.} (a-b) d$I$/d$U$ maps of a vortex at $T=\SI{4.3}{\kelvin}$ and $B=\SI{0}{\milli\tesla}$. (c) Bias spectroscopies far away from the vortex (black squares) and in the centre of the vortex (blue triangles) at $\SI{4.3}{\kelvin}$. The red line shows a fit of Dynes form to the differential conductivity in the superconducting phase with a gap size of $\Delta=\SI{1.24}{\milli e\volt}$.}
    \label{fig:4K_vortex}
\end{figure}

At $\SI{4.3}{\kelvin}$ the stabilization of single vortices within the STM scan frame was substantially harder. Upon repeating the magnetic protocol in small steps seven times, we only found a vortex within our scan frame in a single case. The vortex signature is shown in Fig.~\ref{fig:4K_vortex}. At $U=\SI{1.8}{\milli\volt}\approx \Delta/e$ (a), it appears as a round depression in differential conductance that is slightly smaller than at base temperature which is to be expected due to the temperature dependence of the coherence length $\xi(T)=\xi_0\sqrt{1-T/T_c}$. At zero bias voltage (b), the CdGM states are significantly smeared out. However, a conductance maximum is still found in the centre of the vortex (c) indicating that it is a vortex with odd winding number and, considering its size and shape, most likely $m=1$. A distinction between $\Delta_1$ and $\Delta_2$ is not possible anymore at this temperature due to the temperature broadening of the d$I$/d$U$ spectra (c).

\clearpage

\section*{Supplementary Note 5: Self-consistent calculation of $\Delta$}
\begin{figure}[hpb]
    \centering
    \includegraphics[width=0.75\textwidth]{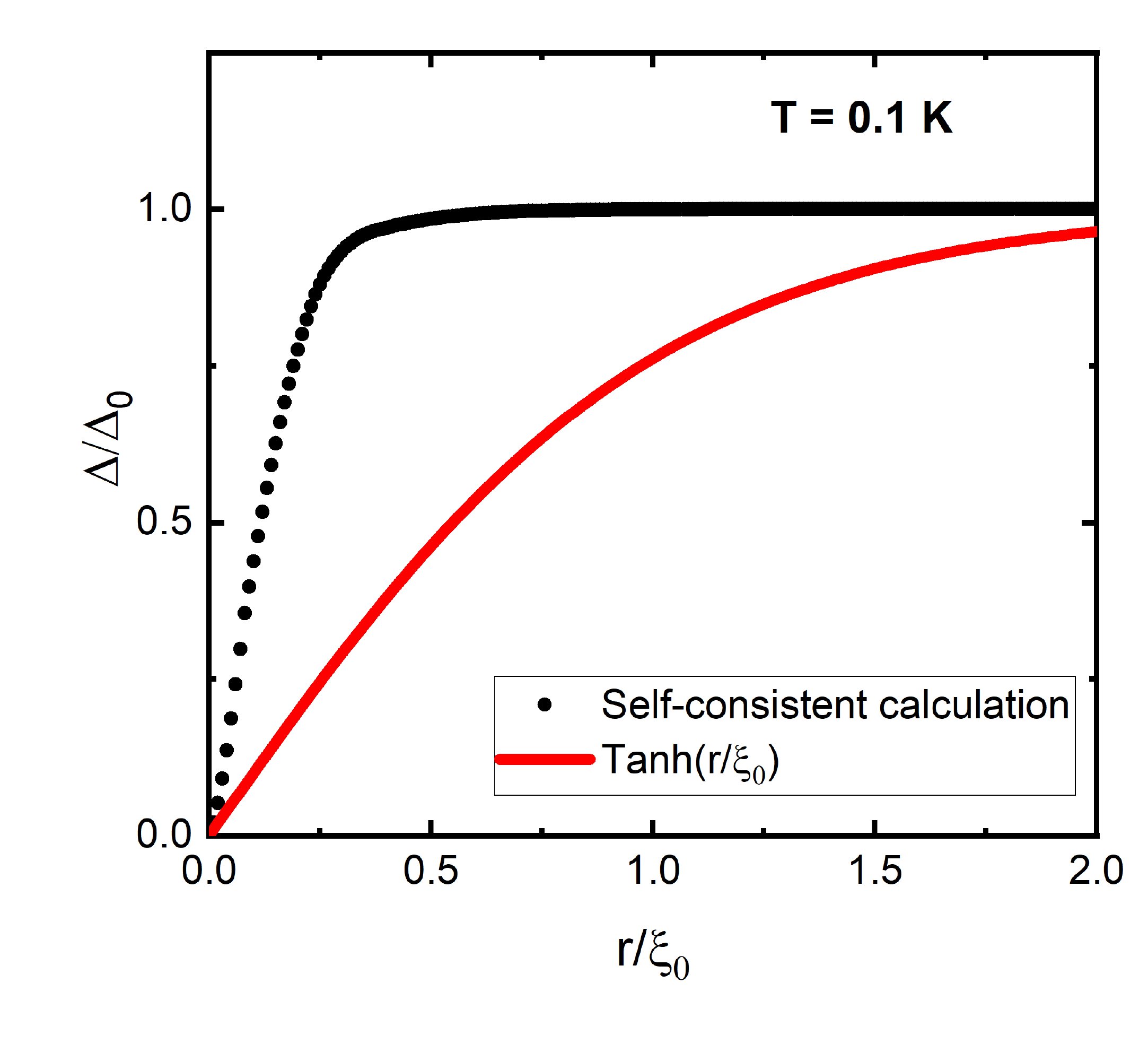}
    \caption{\textbf{Kramer-Pesch effect}: The self-consistent calculation of the pair potential $\Delta$ for a vortex with isotropic Fermi velocity at $T=0.1/7.2\,T_c$ exhibits a shrinking of its core size according to the Kramer-Pesch effect, i.e. a steeper recovery of $\Delta$ close to the vortex centre that does not follow $\tanh{r/\xi_0}$.}
    \label{fig:self-consistent gap}
\end{figure}

We followed Ref. \cite{Silaev2013} and solved the Eilenberger equations in 2D self-consistently for $T=0.1/7.2\,T_c$ in the clean limit, i.e. we refined the pair potential using the self-consistency equation \cite{eilenberger_transformation_1968, Silaev2013}:
\begin{equation}
    \Delta(\bm{r})=2\pi T \Lambda \sum_{n=0}^{N_c} S_F^{-1}\oint_\mathrm{FS}f(i\epsilon_n,\bm{r},\bm{v}_F(\bm{k}))\mathrm{d}^2k
\end{equation}
with 
\begin{equation}
    \Lambda = \left[\log(T/T_c)+\sum\limits_{n=0}^{N_c}\frac{1}{n+1/2}\right]^{-1}
    .
\end{equation}
$S_F$ is the Fermi surface area, $\epsilon_n=2(n+1)\pi T$ the fermionic Matsubara frequencies and $\Lambda$ is the coupling constant. We choose $N_c$ such that $\epsilon_{N_c}=5\,T_c$. Setting the vector potential $\bm{A}=0$ and assuming an isotropic $\bm{v}_F$ we reach convergence 
for $\Delta$ with an accuracy of $10^{-4}\,T_c$. The result for $\Delta(r)$ is displayed in Fig.~\ref{fig:self-consistent gap} along with the function $\Delta(r)=\Delta_0 \tanh{r/\xi_0}$. For a distance from the vortex centre of $r<2\xi_0$ the recovery of $\Delta(r)$ significantly deviates from the function $\Delta_0\tanh{r/\xi_0}$. The slope of $\Delta(r)$ close to $r=0$ is roughly four times as steep as expected from a simple tanh behaviour with universal $\xi_0$. This leads to a core size $\xi^\mathrm{(c)}=\Delta(\infty)\left[\lim_{r\rightarrow 0} \frac{\mathrm{d\Delta(r)}}{\mathrm{d}r}\right]^{-1}$ that is roughly $\xi_0/4$.

\clearpage



\bibliography{supp}